\documentclass[aps,twocolumn,superscriptaddress,altaffilletter,lengthcheck,
tightenlines,showpacs,showkeys]{revtex4}
\usepackage{amsmath} 
\usepackage{amssymb} 
\newcommand{\al}{\alpha}

\newcommand{\ben}{\begin{eqnarray}}
\newcommand{\een}{\end{eqnarray}}
\newcommand{\be}{\begin{equation}}
\newcommand{\ee}{\end{equation}}
\newcommand{\ba}{\begin{eqnarray}}
\newcommand{\ea}{\end{eqnarray}}
\newcommand{\n}{\label}

\newcommand{\la}{\lambda}

\newcommand{\ga}{\gamma}
\newcommand{\ro}{\rho}

\newcommand{\Om}{\Omega}

\newcommand{\bn}{\begin{equation}\label}

\usepackage[dvipdf]{epsfig}
\usepackage{color}

\begin{document}

\title{Dark matter, dark energy, and dark radiation coupled with a transversal interaction}

%%%%%%%%%%%%%%%%%%%%%%%%%%%%%%%%%%%%%%%%%%%%%%%%%%%%%%%%%%%%%%%%%%%%%%%%%%%%%%%%%%%%%%%%%%%%%%%%%%%%
\author{Luis P. Chimento}\email{chimento@df.uba.ar}
\affiliation{Departamento de F\'{\i}sica, Facultad de Ciencias Exactas y Naturales,  Universidad de Buenos Aires, Ciudad Universitaria 1428, Pabell\'on I,  Buenos Aires, Argentina}
\author{Mart\'{\i}n G. Richarte}\email{martin@df.uba.ar}
\affiliation{Departamento de F\'{\i}sica, Facultad de Ciencias Exactas y Naturales,  Universidad de Buenos Aires, Ciudad Universitaria 1428, Pabell\'on I,  Buenos Aires, Argentina}
%%%%%%%%%%%%%%%%%%%%%%%%%%%%%%%%%%%%%%%%%%%%%%%%%%%%%%%%%%%%%%%%%%%%%%%%%%%%%%%%%%%%%%%%%%%%%%%%%%%%

\date{\today}
\bibliographystyle{plain}

\begin{abstract}

We investigate a cosmological scenario with three interacting components that includes dark matter, dark energy, and radiation in the spatially flat Friedmann-Robertson-Walker  universe. We introduce a 3-dimensional internal space, the interaction vector $\mathbf{Q}=(Q_{x}, Q_{m}, Q_{r})$ satisfying the constraint plane $Q_{x}+ Q_{m}+ Q_{r}=0$, the barotropic index vector $\mbox{\boldmath ${\gamma}$}=(\ga_x,\ga_m,\ga_r)$ and select  a transversal interaction vector $\mathbf{Q_t}$ in a sense that $\mathbf{Q_t}\cdot\mbox{\boldmath ${\gamma}$}=0$.  We exactly solve the source equation for a linear $\mathbf{Q_t}$, that depends on the total energy density and its derivatives up to third order, and find all the component energy densities. We obtain a large set of interactions for which the source equation admits a power law solution and show its asymptotic stability by constructing the Lyapunov function. We apply the $\chi^{2}$ method to the observational Hubble data for constraining the cosmic parameters, and analyze the amount of dark energy in the radiation era for the above linear $\mathbf{Q_t}$. It turns to be that our model fulfills the severe bound of $\Omega_{x}(z\simeq 1100)<0.1$  and is consistent with the future constraints achievable by Planck and CMBPol experiments.

\end{abstract} 
\vskip 1cm

\keywords{dark matter, dark energy, dark radiation, linear transversal interaction}
\pacs{}

%\date{\today}
\bibliographystyle{plain}

\maketitle

%\newpage

%%%%%%%%%%%%%%%%%%%%%%%%%%%%%%%%%

\section{Introduction}
%%%%%%%%%%%%%%%%%%%%%%%%%%%%%%%

Based on the large and still growing number of  astronomical observations, one can agree that there is an  exotic component 
called dark energy which represents more than the $70\%$ of the total energy of the Universe.  Its existence has radically changed our standard  paradigm of cosmology mostly because  of its visible effects on the current state of the Universe \cite{Book}.  It turns out that  dark energy is  a repulsive fuel characterized by a strong negative pressure to overcome the slowing down effect of gravity, making the Universe exhibit an accelerated expansion state at the present time \cite{Book}. This tremendous fact has been confirmed by a plethora of observational  tests such as the high redshift Hubble diagram of type Ia supernovae as standard candles \cite{obse1} and accurate measurements of  cosmic microwave background anisotropies \cite{obse2}. According to the current observations, the present-day value of the dark energy density is about 120 orders of magnitude smaller than the energy scales at the end of inflation, so one of the main challenges in the modern cosmology is to understand such deep mismatch. One way to alleviate the aforesaid problem is working within the context of dynamical dark energy models, leaving aside the standard $\Lambda$CDM model. Besides, the necessity of a dark matter component comes from astrophysical evidences of colliding galaxies, gravitational lensing of mass distribution or a  power spectrum  of clustered matter \cite{Book}, \cite{DMobserva}, \cite{dme}.  The first evidence of dark matter's existence  stemmed from the studies performed by  Zwicky in 1934 to the Coma cluster 
of galaxies \cite{dmo} and since its discovery, dark matter has played an essential role for resolving the riddle of the missing mass in the Universe. 
At the present moment, the astrophysical observations from the galactic to the cosmological scales suggest that dark matter is a substantial  
component to the Universe's total matter density \cite{dme} and  sustain that dark matter represents nearly $25\%$ of the total energy matter of the Universe; 
this invisible and nonbaryonic component is the major agent responsible for the large-structure formation in the Universe \cite{Book}, \cite{DMobserva}. Motivated to understand more about the nature of both dark components,  one could consider an exchange of energy between themselves, i.e., the dark matter not only can feel the presence of  the dark energy through a gravitational expansion of the Universe but also can interact between them \cite{jefe1}.  A coupling between dark energy and dark matter changes the background evolution of the dark sector, allowing us to constrain a particular type of interaction and giving  rise to a richer cosmological dynamics compared with noninteracting models \cite{jefe1}. One way to extend the insight about the dark matter-dark energy interacting mechanism is to explore a bigger picture in which a third component is added, perhaps a  weakly interacting radiation term as it occurs within a warm inflation paradigm \cite{WI}. A scenario in which dark energy interacts with both dark matter and radiation was explored in \cite{T1} whereas the validity of the generalized second law of thermodynamics was studied in \cite{T2} without using a particular kind of interaction. Other cases correspond to take the third component as an unparticle fluid \cite{T3a}, unparticle fluid in loop quantum cosmology \cite{T3b}, or  a general unspecified fluid \cite{T4}. As a step forward to constraining dark matter and dark energy with the physic behind  recombination or big-bang nucleosynthesis epochs, a decoupled radiation term was added to the interacting dark sector for taking into account  the stringent bounds  related to  the behavior of dark energy at early times \cite{hmi1}, \cite{hmi2}.

Below we develop a model composed of three interacting fluids and introduce a 3-dimensional internal space where the three interaction terms and barotropic indexes are viewed as vectors in a vector space. We discuss
the existence of a transversal interaction and center our finding in a model with energy exchange proportional to a linear combination of the  total energy density and its derivative up to third order. 
 We also study the stability of a power law solution (scaling solution) with the help of  Lyapunov's theorem. Finally, we perform a cosmological constraint using the Hubble data and the severe bounds for  dark energy at early times. We will use the units $8\pi G=1$ and signature $(-,+,+,+)$ for the metric of the spacetime.

%%%%%%%%%%%%%%%%%%%%%%%%%%%%%%%%%%%%%%%%%%%
\section{The model }
%%%%%%%%%%%%%%%%%%%%%%%%%%%%%%%%%%%%%%%%%%%

We consider a spatially flat homogeneous  and isotropic Universe described by Friedmann-Robertson-Walker (FRW) spacetime with a line element given by  $ds^{2}=-dt^{2}+a^{2}(t) (dx^{2}+dy^{2}+dz^{2})$ being  $a(t)$ the scale factor. The Universe is  filled with three interacting fluids, namely, dark energy, dark matter, and radiation so that  the evolution of the FRW Universe is governed by  the Friedmann and conservation equations, respectively,  
\be
\n{01}
3H^{2}=\ro= \ro_{x}+\ro_{m}+\ro_{r},
\ee
\be
\n{02}
\dot{\ro}+3H(\ro_{x}+p_{x}+\ro_{m}+p_{m}+\ro_{r}+p_{r})=0,
\ee
where $H = \dot a/a$ is the Hubble expansion rate. Introducing the variable $\eta = \ln(a/a_0)^{3}$, with $a_0$ the present value of the scale factor and $' \equiv d/d\eta$, Eq. (\ref{02}) can be recast 
\be
\n{03}
\ro'=-\gamma_{x}\ro_{x}-\gamma_{m}\ro_{m}- \gamma_{r}\ro_{r},
\ee
where $\gamma_{i}=1+p_{i}/\ro_{i}$  is the barotropic index of each component with $i=\{x,m,r\}$ and $0<\ga_x<\ga_m<\ga_r$.  The interaction terms $3H Q_{i}$ between the components are introduced by  splitting (\ref{03}) into three equations:
\be
\n{04}
\ro_x' + \ga_{x} \ro_x =  Q_{x}.
\ee
\be
\n{05}
\ro_m' + \ga_{m} \ro_m = Q_{m},
\ee
\be
\n{06}
\ro_r' + \ga_{r} \ro_r = Q_{r},
\ee
where the $Q_{i}$ describe the energy transfer and satisfy the condition
\be
\n{06b}
 Q_{x}+  Q_{m}+ Q_{r}=0,
\ee
to recover the whole conservation equation (\ref{03}) after having summed  Eqs.  (\ref{04})-(\ref{06}).
Also, we assume that all component energy densities are definite positive.

To investigate the proposed model, we introduce a 3-dimensional internal space with an orthonormal vector basis $\{\mathbf{e_t},\mathbf{e_o},\mathbf{n}\}$ defined in the following way: we set the coordinate origin at the intersection of the interaction plane (\ref{03}) with the vector formed with the barotropic indexes $\mbox{\boldmath ${\gamma}$}=(\ga_x,\ga_m,\ga_r)$, the pair of vectors $\{\mathbf{e_t},\mathbf{e_o}\}$ is contained in the interaction plane, $\mathbf{e_t}$ is orthogonal to the vector $\mbox{\boldmath ${\gamma}$}$, so $\mbox{\boldmath ${\gamma}$}\cdot\mathbf{e_t}=0$, the orthogonal projection of the vector $\mbox{\boldmath ${\gamma}$}$ on the interaction plane defines the direction of the vector $\mathbf{e_o}$ and the normal vector $\mathbf{n}=(1,1,1)/\sqrt{3}$ is orthogonal to the interaction plane. The interaction terms $Q_{i}$ are also viewed as the components of a vector $\mathbf{Q}=(Q_{x}, Q_{m}, Q_{r})$ that lives on the plane $ \Pi:Q_{x}+Q_{m}+Q_{r}=0$, meaning that
\bn{vQ}
\mathbf{Q}=q_t\,\mathbf{e_t}+q_o\,\mathbf{e_o},
\ee
where $q_t$ and $q_o$ are the components of the interaction vector $\mathbf{Q}$ on the plane $\Pi$ and $\mathbf{n}\cdot \mathbf{Q}=0$.

Taking into account that $\mathbf{e_t}$ is the unique vector of the basis with the property of being orthogonal to $\mbox{\boldmath ${\gamma}$}$, we adopt this property as a simple criteria for selecting only those interactions which are collinear with the aforesaid preferred direction in the plane $\Pi$ that we call ``transversal interaction'', so 
\bn{qt}
\mathbf{Q_t}=q_t\,\mathbf{e_t}
\ee
with $Q_t=q_t$  ensuring that we can always  take $\mbox{\boldmath ${\gamma}$}\cdot \mathbf{Q}=0$ uniquely. The transversal character of the interaction vector (\ref{qt}) will simplify enough the equations which determine the component energy densities as we  will see below. 

\begin{figure}[h!]
\begin{center}
\includegraphics[height=6.9cm,width=7.5cm]{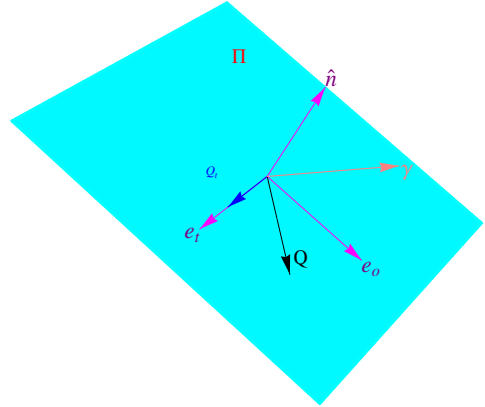}
\caption{The plot shows the orthonormal vector basis formed by $\{\mathbf{e_t},\mathbf{e_o},\mathbf{n}\}$, the interaction plane $\Pi: Q_{x}+Q_{m}+Q_{r}=0$, the interaction vector  $\mathbf{Q}=q_t\,\mathbf{e_t}+q_o\,\mathbf{e_o}$ and the vector $\mbox{\boldmath ${\gamma}$}=\ga_n\mathbf{n}+\ga_o\mathbf{e_o}$.}
\label{F1}
\end{center}
\end{figure} 
%\vspace{1cm}

The two basis vectors that span the interaction plane $\Pi$ are given by
\be
\n{v1}
 \mathbf{e_t}=\frac{(\ga_{m}-\ga_{r}, \ga_{r}-\ga_{x}, \ga_{x}-\ga_{m})}{e},
\ee
\be
\n{v2}
 \mathbf{e_o}=\frac{(\ga_{m}+\ga_{r}-2\ga_{x}, \ga_{x}+\ga_{r}-2\ga_{m}, \ga_{x}+\ga_{m}-2\ga_{3})}{\sqrt{3}\,e},
\ee
where $e^2= 3\sum_{i}\ga^{2}_{i}-[\sum_{i}\ga_{i}]^{2}$  while the vector built with the barotropic indexes $\mbox{\boldmath ${\gamma}$}=\ga_n\mathbf{n}+\ga_o\mathbf{e_o}$ lives in the plane spanned by the normal  $\mathbf{n}$ and $\mathbf{e_o}$ [see Fig.(\ref{F1})]. 

Now, we will construct an interacting three fluid model with the transversal interaction (\ref{qt}). After differentiating Eq. (\ref{03}) and using Eqs. (\ref{04})-(\ref{06}) we have   
\be
\n{07}
\ro''=\ga^{2}_{x}\ro_{x}+\ga^{2}_{m}\ro_{m}+ \ga^{2}_{r}\ro_{r}.
\ee
Solving the algebraic system of equations in the $(\ro_{x}, \ro_{m}, \ro_{r})$ variables (\ref{01}), (\ref{03}), and (\ref{07}), we obtain $\ro_{x}$,  $\ro_{m}$, and  $\ro_{r}$ as a function of $\ro$, $\ro'$, and $\ro''$ only;
\be
\n{08}
\ro_x=\frac{\ga_{m}-\ga_{r}}{\Delta}\left[ \ga_{m}\ga_{r}\ro+(\ga_{m}+\ga_{r})\ro'+ \ro''\right],
\ee
\be
\n{09}
\ro_m=-\frac{\ga_{x}-\ga_{r}}{\Delta}\left[ \ga_{x}\ga_{r}\ro+ (\ga_{x}+\ga_{r})\ro'+ \ro''\right],
\ee
\be
\n{10}
\ro_r=\frac{\ga_{x}-\ga_{m}}{\Delta}\left[ \ga_{x}\ga_{m}\ro+ (\ga_{x}+\ga_{m})\ro'+\ro''\right],
\ee
where $\Delta=(\ga_{x}-\ga_{m}).(\ga_{x}-\ga_{r}).(\ga_{m}-\ga_{r})$ is the determinant of that algebraic system of equations.  Equations (\ref{08}), (\ref{09}), and (\ref{10}) clearly represent the straightforward extension of the case studied in \cite{jefe1}  where  an interacting two-fluid scenario for the dark sector in the FRW Universe was investigated. Following Ref. \cite{jefe1},  we replace (\ref{08}) into (\ref{04}), (\ref{09}) into (\ref{05}), or (\ref{10}) into (\ref{06}) and obtain the same third order differential equation, that we call ``source equation'', for the total energy density; 
$$
\ro'''+(\ga_{x}+\ga_{m}+\ga_{r})\ro''+
$$
\be
\n{11}
(\ga_{x}\ga_{r}+\ga_{x}\ga_{m}+\ga_{m}\ga_{r})\ro'+\ga_{x}\ga_{m}\ga_{r}\ro={\cal Q},
\ee
where its source term ${\cal Q}$ involves a linear combination of the interaction vector components  $Q_{i}$    
\be
\n{12}
{\cal Q}=\ga_{x}Q_{m}\ga_{r}+ \ga_{r}Q_{x}\ga_{m}+\ga_{m}Q_{r}\ga_{x}.
\ee
By combining the transversal interaction (\ref{qt}), the basis vector (\ref{v1}) and  Eq. (\ref{12}) we find
\bn{qs}
{\cal Q}=-\Delta \frac{q_t}{e}
\ee

Finally, once the transversal interaction $\mathbf{Q_t}$ is specified we obtain the energy density $\ro$ by solving  the source equation Eq. (\ref{11}) and the component energy densities $\ro_x$, $\ro_m$, and $\ro_r$ after inserting $\ro$ into Eqs. (\ref{08}), (\ref{09}), and (\ref{10}). 

%%%%%%%%%%%%%%%%%%%%%%%%%%%%%%%%%%%%%%%%%%%%%%%%%%%%%%
\subsection{Stability analysis}
%%%%%%%%%%%%%%%%%%%%%%%%%%%%%%%%%%%%%%%%%%%%%%%%%%%%%

Now, we will investigate the stability of power law solutions. To this end, we will assume that the source equation (\ref{11}) admits scaling solutions and subsequently we will look for the set of interaction terms that give rises to them. The knowledge of power law solutions is very useful because it determines the asymptotic behavior of the effective barotropic index $\ga=(\ga_x\ro_x+\ga_m\ro_m+\ga_r\ro_r)/(\ro_x+\ro_m+\ro_r)=-2\dot{H}/3H^{2}$, which ranges between $\ga_x<\ga<\ga_r$. These solutions represent a Universe approaching to a stationary stage characterized by $\ga=\ga_s$ and $a=t^{2/3\ga_{s}}$; then the existence of the attractor solution $\ga_{s}$ will imply that $\ga$ goes to the asymptotic constant value $\ga\to\ga_s$. So on the attractor 
\be
\n{13}
\ga_{s}=\frac{\ga_x\ro_{xs}+\ga_m\ro_{ms}+\ga_r\ro_{rs}}{\ro_{xs}+\ro_{ms}+\ro_{rs}}=\frac{\ga_{x}+\ga_{m}r_{mx}+\ga_{r}r_{rx}}{1+r_{mx}+r_{rx}},   
\ee
or
\be
\n{13b}
(\ga_{x}-\ga_{s})+ (\ga_{m}-\ga_{s})r_{mx}+(\ga_{r}-\ga_{s})r_{rx}=0,  
\ee
where we have defined the ratios $r_{mx}=\ro_{ms}/\ro_{xs}$ and $r_{rx}=\ro_{rs}/\ro_{xs}$. From the positivity of the component energy densities, the small value of the ratio $r_{rs}=\Om_{rs}/\Om_{xs}$, the range of the effective barotropic index $\ga_x<\ga<\ga_r$, and tEq. (\ref{13b}), we determine that $\ga_s$ ranges between $\ga_{x}<\ga_{s}<\ga_{m}<\ga_{r}$ and the ratios $r_{mx}$ and $r_{rx}$ become asymptotically constant on the attractor stage, alleviating the cosmic coincidence problem. In the case of $\ga_{s}=0$,  we have a final de Sitter  regime, $H=cte$ with  $r_{mx}=-(\ga_{x}/\ga_{m})-r_{rx}(\ga_{r}/\ga_{m})$. 

To investigate the existence of a power law attractor solution $a=t^{2/3\ga_s}$, we use that $\ro'=-\ga\ro$, $\ro''=(\ga^{2}-\ga')\ro$, $\ro'''=-(\ga^{3}-3\ga\ga'+ \ga'')\ro$, change the source equation (\ref{11}) into
\be
\n{14a}
\ga''+(\ga_{x}+\ga_{m}+\ga_{r}-3\ga)\ga'- {\cal P}(\ga)=-\frac{{\cal Q}}{\ro},
\ee
\be
\n{14b}
{\cal P}(\ga)=-(\ga-\ga_{x})(\ga-\ga_{m})(\ga-\ga_{r}),
\ee
and impose both (i) that $\ga_{s}$ be a constant stationary solution of Eq. (\ref{14a}) and (ii) the stability condition so that $\ga_{s}$ is stable. The existence of $\ga_{s}$ implies $\ga_{s}'=0$, $\ga_{s}''=0$ and  
\be
\n{15}
{\cal Q}(\ga_{s})={\cal P}(\ga_{s})\,\ro, \qquad   {\cal Q}(\ga_{s})<0,
\ee
due to Eq. (\ref{13b}).

Taking into account the specific form of the interaction (\ref{15}), we will assume separability and the stability analysis will be performed for interactions of the form 
\be
\n{16}
{\cal Q}(\ro, \ga, \ga', \ga'')={\cal P}(\ga){\cal K}(\ga, \ga', \ga'')\,\ro
\ee
Combining Eqs. (\ref{14a}) and (\ref{16}), we write the equation governing  the dynamical evolution of the barotropic index in a simpler form
\be
\n{17}
\ga''+(\ga_{x}+\ga_{m}+\ga_{r}-3\ga)\ga'= -{\cal P}(\ga)[{\cal K}-1],
\ee
where the function ${\cal K}$ fulfills the condition,
\be
\n{18a}
{\cal K}(\ga=\ga_{s}, \ga'=0, \ga''=0)=1,
\ee
for assuring the existence of the constant stationary solutions $\ga_{s}$. Perturbing around the solution $\ga_{s}$ by taking $\ga=\ga_{s}+\epsilon$ with $|\epsilon/\ga_s|\ll 1$, Eq. (\ref{17}) can be recast as
\be
\n{18b}
\epsilon''+ (\ga_{x}+\ga_{m}+\ga_{r}-3\ga_{s})\epsilon'=-{\cal P}(\ga_{s})[{\cal K}_{\ga} \epsilon+{\cal K}_{\ga'} \epsilon'+ {\cal K}_{\ga''} \epsilon''],
\ee
where  ${\cal K}_{\ga}$, ${\cal K}_{\ga'}$, and  ${\cal K}_{\ga''}$ stand for the partial derivatives with respect to $\ga$, $\ga'$, and $\ga''$, respectively. These derivatives   are evaluated at the point $(\ga=\ga_{s},\ga_{s}'=0, \ga_{s}''=0)$ and we have used that ${\cal K}(\ga=\ga_{s}+\epsilon, \epsilon')=1+{\cal K}_{\ga}\epsilon+{\cal K}_{\ga'}\epsilon'+{\cal K}_{\ga''}\epsilon''+{\cal O}(\epsilon^2, \epsilon'^2, \epsilon''^2)$. It turns out  that (\ref{18b}) can be written as  the equation of motion for a dissipative or antidissipative mechanical system. This resemblance emerges from the analogy with the classical potential problem
\be
\n{19}
\frac{d}{d \eta}\left[\frac{\epsilon'^{2}}{2} +{\cal  V}(\epsilon)\right]=-\alpha\epsilon'^{2},
\ee
where $\alpha=[\ga_{x}+\ga_{m}+\ga_{r}-3\ga_{s}+{\cal P}(\ga_{s}) {\cal K}_{\ga'}]/[1+{\cal P}(\ga_{s}) {\cal K}_{\ga''}]$  and the potential 
\be
\n{19b}
{\cal  V}(\epsilon)=\beta\frac{\epsilon^{2}}{2},
\ee
with $\beta={\cal P}(\ga_{s}) {\cal K}_{\ga}/[1+{\cal P}(\ga_{s}) {\cal K}_{\ga''}]$. In order to assure the stability of the scaling solution $\ga=\ga_{s}$, we demand that  both coefficients $\al$ and $\beta$ are positives. So, for any transversal interaction $\mathbf{Q_t}$ leading to $\al>0$, the potential  ${\cal  V}$ has a minimum at $\epsilon=0$ when $\beta>0$, the function inside the  square bracket in Eq. (\ref{19}) is a Lyapunov function, and the perturbation decreases asymptotically reaching  $\epsilon=0$ (attractor) in the limit $\eta \rightarrow \infty$; then the system is asymptotically stable in the sense of Lyapunov. In other words,  when the condition (\ref{18a}) is fulfilled $\ga_{s}$ becomes a constant stationary solution of Eq. (\ref{17}) and it is stable whenever the stability conditions $\al>0$ and $\beta>0$ are satisfied. Note that the interaction (\ref{16}) depends complicatedly on $\ro$, $\ro'$, $\ro''$ and $\ro'''$ because the term ${\cal K}(\ga, \ga', \ga'')$ depends on $\ga=-\ro'/\ro$, $\ga'=(\ro'/\ro)^{2}-\ro''/\ro$, and $\ga''=-(\ro'''/\ro)-(\ro'/\ro)^{3}+3(\ro'/\ro)[\ro''/\ro-(\ro'/\ro)^{2}]$; then ${\cal K}$ contains  some nontrivial terms implying that our findings are indeed valid even for a broad set of nonlinear interactions. 

%%%%%%%%%%%%%%%%%%%%%%%%%%%%%%%%%%%%%%%%%%%%%%%%%%%%%%%%%%%%%%%%%%%%%%%%%%%%%%%%
\subsection{Linear  transversal  interaction $\mathbf{Q_t}$ }
%%%%%%%%%%%%%%%%%%%%%%%%%%%%%%%%%%%%%%%%%%%%%%%%%%%%%%%%%%%%%%%%%%%%%%%%%%%%%%%%

Following \cite{jefe1}, we assume a transversal interaction $\mathbf{Q_t}$, with $\mbox{\boldmath ${\gamma}$}\cdot \mathbf{Q_t}=0$, which is a linear combination of $\ro_{x}$, $\ro_{c}$, $\ro_{r}$, their derivatives up to first order,  $\ro$, $\ro'$, $\ro''$, and $\ro'''$, so from Eq. (\ref{qs}) we have
\[{\cal Q}=\al_{1}\ro_{x}+\al_{2}\ro_{m}+\al_{3}\ro_{r}+\al_{4}\ro'  \]
\be
\n{GG}
+\al_{5}\ro''+\al_{7}\ro'''+\al_{8}\ro'_{x}+\al_{9}\ro'_{m}+\al_{10}\ro'_{r}.
\ee
Using Eqs. (\ref{08}), (\ref{09}), and (\ref{10}), we can recast (\ref{GG}) as
\be
\n{GG2}
{\cal Q}=\beta_{1}\ro+\beta_{2}\ro'+\beta_{3}\ro''+\beta_{4}\ro'''.
\ee
where the new $\beta_i$ coefficients are written in terms of the old $\alpha_i$ ones.  Eq. (\ref{GG2}) clearly shows that the most general linear interaction only requires a linear combination of $\ro$ and its derivatives up to third order.  Combining Eqs. (\ref{16}) and (\ref{GG2}) along with $\ro'=-\ga\ro$, $\ro''=(\ga^{2}-\ga')\ro$, and  $\ro'''=-(\ga^{3}-3\ga\ga'+ \ga'')\ro$, we obtain that 
\be
\n{K}
{\cal K}=\frac{\beta_{1}-\beta_{4}(\ga^{3}+\ga''+3\ga\ga')+\beta_{3}\ga^{2}-\beta_{3}\ga'-\beta_{2}\ga}{{\cal P}(\ga)}.
\ee
Applying the condition (\ref{18a}) referring to the existence of the constant solution $\ga_{s}$ to (\ref{K}), we get a constraint
\be
\n{K2}
\beta_{1}-\beta_{4}\ga^{3}_{s}+\beta_{3}\ga^{2}_{s}-\beta_{2}\ga_{s}={\cal P}(\ga_{s}),
\ee
for the coefficients $\beta_{1}$, $\beta_{2}$, $\beta_{3}$,  and $\beta_{4}$. By solving Eq.(\ref{K2}) for $\beta_{2}$ and inserting into Eq. (\ref{GG2}), we obtain the final form of the effective interaction term
\be
\n{GLI}
{\cal Q}=\beta_{1}\ro+\ga^{-1}_{s}[\beta_{1}-{\cal P}(\ga_{s}) -\beta_{4}\ga^{3}_{s}+\beta_{3}\ga^{2}_{s}]\ro'+\beta_{3}\ro''+\beta_{4}\ro''',
\ee
so in the end, the most  general linear  interaction (\ref{GG}) or (\ref{GLI}) only involves four parameters, namely,  $\beta_{1}$, $\beta_{3}$, $\beta_{4}$, and $\ga_{s}$. Replacing (\ref{GLI}) into Eq. (\ref{14a}), we find the polynomial $\ga^{3}+A\ga^{2}+B\ga+C=0$ for constant $\ga$ solutions;  one of its roots is  $\ga_{s}$ ,  whereas the other two are given by
\be
\n{Ra1}  
\ga_{-}=\left(\frac{A+\ga_{s}}{2}\right)\left[-1 \pm \sqrt{1+\frac{4C}{\ga_{s}(A+\ga_{s})^{2}}}~\right],  
\ee
\be
\n{Ra2}  
\ga_{+}=-\frac{C}{\ga_{-}\ga_{s}} 
\ee
where the coefficients $A$, $B$, and $C$ turn out  to be 
\be
\n{R1}  
A=\frac{\sum_{i}\ga_{i}-\beta_{3}}{\beta_{4}-1}, \qquad C=\frac{\ga_{x}\ga_{m}\ga_{r}-\beta_{1}}{\beta_{4}-1},
\ee
\be
\n{R2}  
B=\frac{\sum_{i\neq j}{}\ga_{i}\ga_{j}-\ga^{-1}_{s}[\beta_{1}-{\cal P}(\ga_{s}) -\beta_{4}\ga^{3}_{s}+\beta_{3}\ga^{2}_{s}]}{\beta_{4}-1}. 
\ee
The exact solution of  the source equation (\ref{11}) for the general linear transversal interaction (\ref{GLI}) is given by 
\be
\n{den1}
\ro=b_{1}a^{-3\ga_{s}}+b_{2}a^{-3\ga_{+}}+b_{3}a^{-3\ga_{-}}
\ee
whereas the  component energy densities are obtained from Eqs. (\ref{08}), (\ref{09}), and (\ref{10}):
\[\ro_x=\frac{\ga_{m}-\ga_{r}}{\Delta}[ \ga_{m}\ga_{r}(b_{1}a^{-3\ga_{s}}+ b_{2}a^{-3\ga_{+}}+ b_{3} a^{-3\ga_{-}})\]
\[-(\ga_{m}+\ga_{r})(b_{1}\ga_{s}a^{-3\la_{s}}+ b_{2}\ga_{+}a^{-3\ga_{+}}+ b_{3}\ga_{-} a^{-3\ga_{-}})\]
\be
\n{08b}
+ (\ga^{2}_{s}b_{1} a^{-3\ga_{s}}+ \ga^{2}_{+}b_{2} a^{-3\ga_{+}}+\ga^{2}_{-}b_{3} a^{-3\ga_{-}})],
\ee
\[\ro_m=-\frac{\ga_{x}-\ga_{r}}{\Delta}[ \ga_{x}\ga_{r}(b_{1}a^{-3\ga_{s}}+ b_{2}a^{-3\ga_{+}}+ b_{3} a^{-3\ga_{-}})\]
\[-(\ga_{x}+\ga_{r})(b_{1}\ga_{s}a^{-3\ga_{s}}+ b_{2}\ga_{+}a^{-3\ga_{+}}+ b_{3}\ga_{-} a^{-3\ga_{-}})\]
\be
\n{09b}
+ (\ga^{2}_{s}b_{1} a^{-3\ga_{s}}+ \ga^{2}_{+}b_{2} a^{-3\ga_{+}}+\ga^{2}_{-}b_{3} a^{-3\ga_{-}})],
\ee
\[\ro_r=\frac{\ga_{x}-\ga_{m}}{\Delta}[ \ga_{m}\ga_{x}(b_{1}a^{-3\ga_{s}}+ b_{2}a^{-3\ga_{+}}+ b_{3} a^{-3\ga_{-}})\]
\[-(\ga_{x}+\ga_{m})(b_{1}\ga_{s}a^{-3\ga_{s}}+ b_{2}\ga_{+}a^{-3\ga_{+}}+ b_{3}\ga_{-} a^{-3\ga_{-}})\]
\be
\n{10b}
+ (\ga^{2}_{s}b_{1} a^{-3\ga_{s}}+ \ga^{2}_{+}b_{2} a^{-3\ga_{+}}+\ga^{2}_{-}b_{3} a^{-3\ga_{-}})],
\ee
In general, the total energy density in terms of the physical quantities such as density parameters is given by
\be
\n{DenZ}
\ro=3H^{2}_{0}\Big({\cal A}x^{3\ga_{s}}+{\cal B}x^{3\ga_{+}}+{\cal C}x^{3\ga_{-}}\Big)
\ee
with $x=z+1$ and the  $z$  cosmological redshift while the integration constants are given by
\[{\cal A}=\frac{\Omega_{m0}(\ga_{+}-\ga_{m})(\ga_{-}-\ga_{m})+\Omega_{r0}(\ga_{+}-\ga_{r})(\ga_{-}-\ga_{r})}{(\ga_{s}-\ga_{+})(\ga_{s}-\ga_{-})}\]
\be
\n{A}
+\frac{\Omega_{x0}(\ga_{+}-\ga_{x})(\ga_{-}-\ga_{x})}{(\ga_{s}-\ga_{+})(\ga_{s}-\ga_{-})}
\ee
\[{\cal B}=\frac{\Omega_{m0}(\ga_{s}-\ga_{m})(\ga_{m}-\ga_{-})+\Omega_{0r}(\ga_{s}-\ga_{r})(\ga_{r}-\ga_{-})}{(\ga_{s}-\ga_{+})(\ga_{+}-\ga_{-})}\]
\be
\n{B}
+\frac{\Omega_{x0}(\ga_{s}-\ga_{x})(\ga_{x}-\ga_{-})}{(\ga_{s}-\ga_{+})(\ga_{+}-\ga_{-})}
\ee
\[{\cal C}=\frac{\Omega_{m0}(\ga_{s}-\ga_{m})(\ga_{m}-\ga_{+})+\Omega_{r0}(\ga_{s}-\ga_{r})(\ga_{r}-\ga_{+})}{(\ga_{s}-\ga_{-})(\ga_{-}-\ga_{+})}\]
\be
\n{C}
+\frac{\Omega_{x0}(\ga_{s}-\ga_{x})(\ga_{x}-\ga_{+})}{(\ga_{s}-\ga_{-})(\ga_{-}-\ga_{+})}
\ee
where $\Omega_{0i}=\ro_{0i}/3H^{2}_{0}$ are  density parameters  fulfilling the condition $\Omega_{x0}+\Omega_{r0}+\Omega_{m0}=1$ for a spatially flat FRW Universe. Additionally, we will choose  $(\ga_{x}, \ga_{m}, \ga_{r})=(0,1,4/3)$ to recover the three self-conserved cosmic components in the limit of vanishing interaction. 

%%%%%%%%%%%%%%%%%%%%%%%%%%%%%%%%%%%%%%%%%%%%%%%%%%%%%%%%%%%%%%%%%%%%%%%%%%%%%%%%
\subsection{Observational constraints on a transversal interacting  model}
%%%%%%%%%%%%%%%%%%%%%%%%%%%%%%%%%%%%%%%%%%%%%%%%%%%%%%%%%%%%%%%%%%%%%%%%%%%%%%%%
We will provide a qualitative estimation of the cosmological paramaters by constraining them with the Hubble data  \cite{obs3}- \cite{obs4} and the strict bounds for the behavior of dark energy at early times \cite{EDE1}-\cite{EDE2}. In the former case, the  statistical analysis is based on the $\chi^{2}$ function of the Hubble data which is constructed as (e.g., \cite{Press})
\be
\n{c1}
\chi^2(\theta) =\sum_{k=1}^{12}\frac{[H(\theta,z_k) - H_{obs}(z_k)]^2}{\sigma(z_k)^2},
\ee
where $\theta$ stands for cosmological parameters, $H_{obs}(z_k)$ is the observational $H(z)$ data at the redshift $z_k$, $\sigma(z_k)$ is the corresponding $1\sigma$ uncertainty, and the summation is over the $12~$ observational  $H(z)$ data. The Hubble function is not integrated over and it is directly related with the properties of the dark energy, since its value comes from the cosmological observations. Using the absolute ages of passively evolving galaxies observed at different redshifts, one obtains the differential ages $dz/dt$ and the function $H(z)$ can be measured through the relation $H(z)=-(1+z)^{-1}dz/dt$ \cite{obs3}, \cite{obs4}. The data  $H_{obs}(z_i)$ and $H_{obs}(z_k)$ are uncorrelated because they were obtained from the observations of galaxies at different redshifts. 

From Eq. (\ref{DenZ}), one finds that the Hubble expansion of the model  becomes 
\be
\n{Ht}
H(\theta| z)=H_{0} \Big( {\cal A}x^{3\ga_{s}}+{\cal B}x^{4}+{\cal C}x^{3}\Big)^{\frac{1}{2}}
\ee
where ${\cal A}$,  ${\cal B}$, and  ${\cal C}$ are  obtained form (\ref{A}), (\ref{B}), and (\ref{C}), respectively. For practical reasons  mentioned in the last section we have simply selected $\ga_{+}=4/3$, $\ga_{-}=1$.

Here, we consider  $\theta=\{H_{0},\ga_{s},  \Omega_{x0},\Omega_{m0}\}$ plus the constraint on the density parameters to assure the flatness condition $(\Omega_{r0}=1-\Omega_{x0}-\Omega_{m0})$; then  we have four independent parameters only. We will take  two independent parameters and will  fix the other ones along the statistic analysis until all parameters have been varied  and estimated with the $\chi^{2}$ function. Then, for a given pair of $(\theta_{1}, \theta_{2})$,  we are going to perform the statistical analysis by minimizing the $\chi^2$ function to  obtain the best-fit values of  the random variables $\theta_{c}=\{\theta_{1c}, \theta_{2c} \}$ that correspond to a maximum of Eq.(\ref{c1}). More precisely, the  best-fit parameters $\theta_{c}$ are those values where $\chi^2_{min}(\theta_{c})$ leads to the local minimum of the $\chi^2(\theta)$ distribution. If $\chi^2_{d.o.f.}=\chi^2_{min}(\theta_{c})/(N -n) \leq 1$ the fit is good and the data are consistent with the considered model $H(z|\theta)$. Here, $N$ is the number of data and $n$ is the number of parameters \cite{Press}. The variable $\chi^2$ is a random variable that depends on $N$ and its probability distribution is a $\chi^2$ distribution for $N-n$ degrees of freedom. Besides, $68.3\%$ confidence  contours  in the 2D plane  are made of the random data sets that satisfy the inequality $\Delta\chi^{2}=\chi^2(\theta)-\chi^{2}_{min}(\theta_{c})\leq 2.30$. The latter equation defines a bounded region by a closed area around $\theta_{c}$ in the two-dimensional parameter plane; thus,  the $1\sigma$ error bar can be identified with the distance from the $\theta_{c}$ point to the boundary of the  two-dimensional parameter plane. It can be shown that $95.4\%$ confidence contours  with a $2\sigma$  error bar in the samples satisfy $\Delta\chi^{2}\leq 6.17$.

\begin{figure}[hbt!]
\begin{minipage}{1\linewidth}
\resizebox{1.6in}{!}{\includegraphics{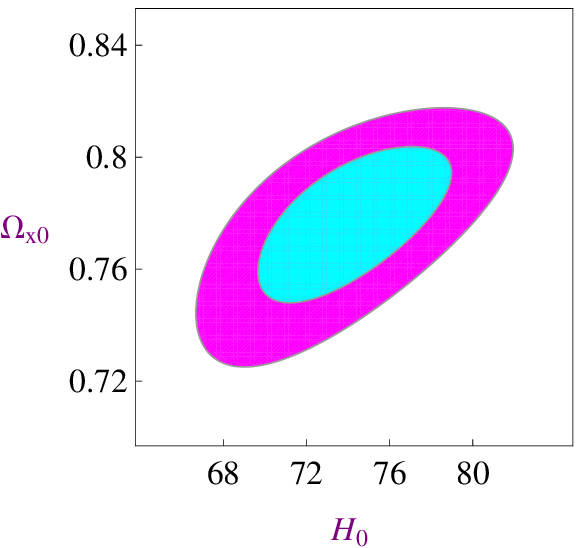}}
\resizebox{1.6in}{!}{\includegraphics{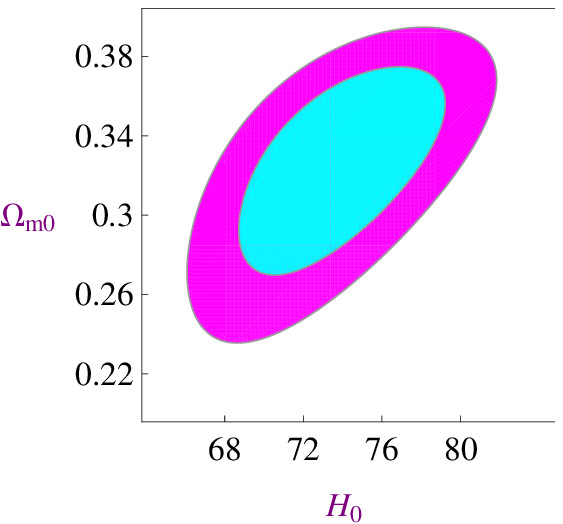}}\hskip0.05cm
\resizebox{1.6in}{!}{\includegraphics{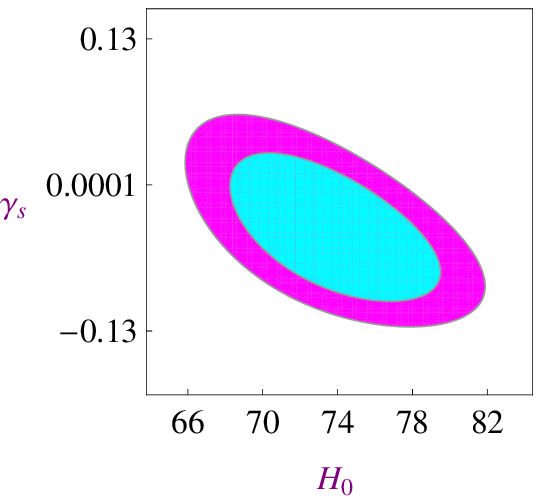}}\hskip0.05cm 
\resizebox{1.6in}{!}{\includegraphics{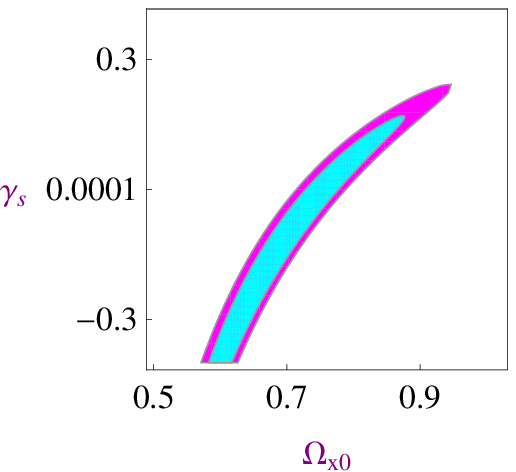}}\hskip0.05cm 
\resizebox{1.6in}{!}{\includegraphics{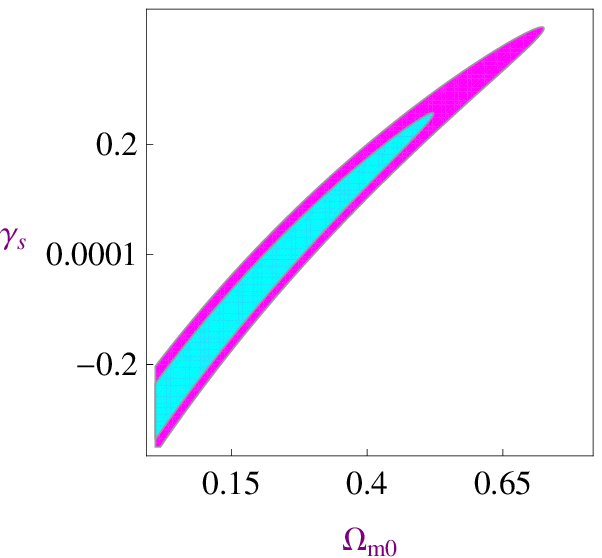}}
\resizebox{1.6in}{!}{\includegraphics{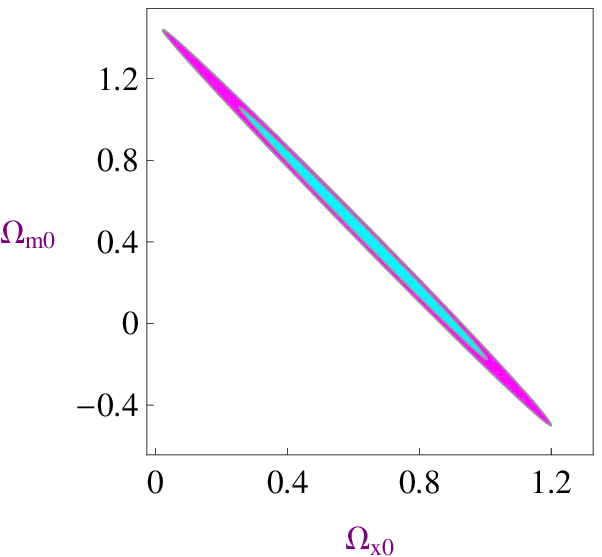}}
\caption{\scriptsize{Two-dimensional C.L. associated with $1\sigma$,$2\sigma$ for different $\theta$ planes.}}
\label{Fig1}
\end{minipage}
\end{figure}

\begin{figure}[hbt!]
\begin{minipage}{1\linewidth}
\resizebox{1.6in}{!}{\includegraphics{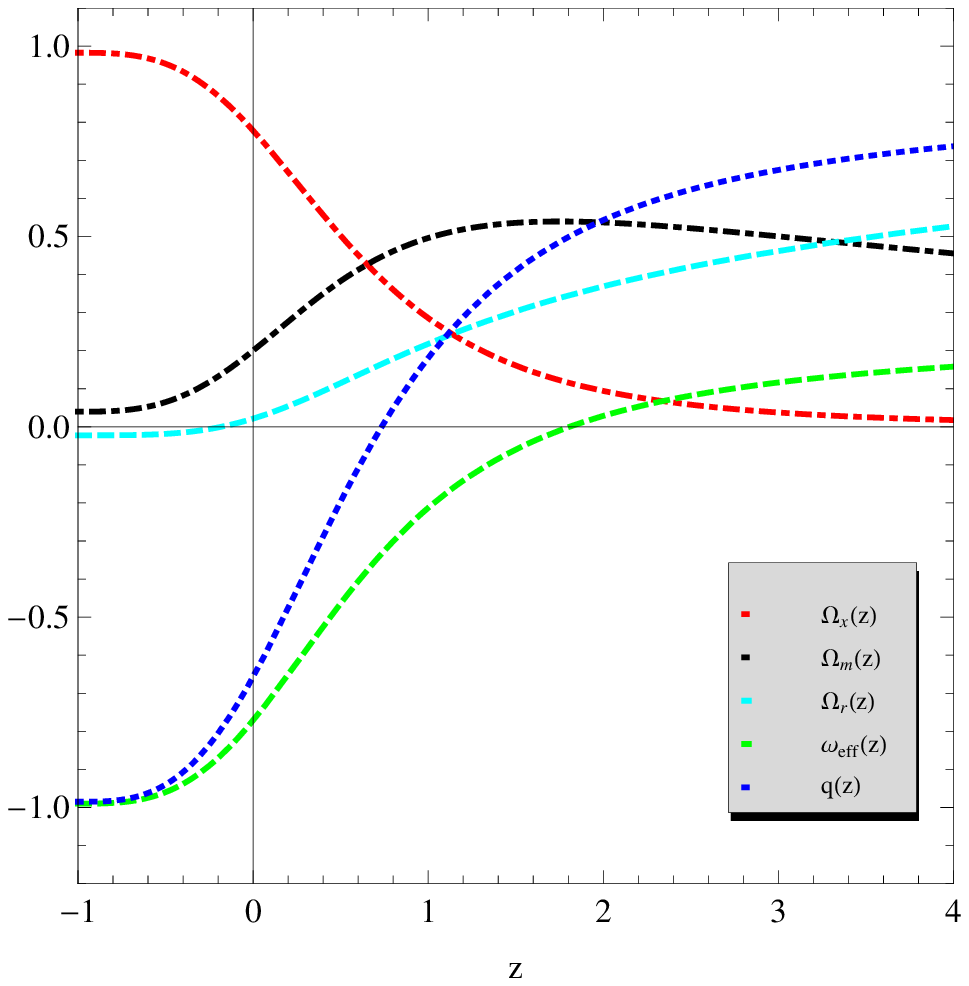}}
\resizebox{1.6in}{!}{\includegraphics{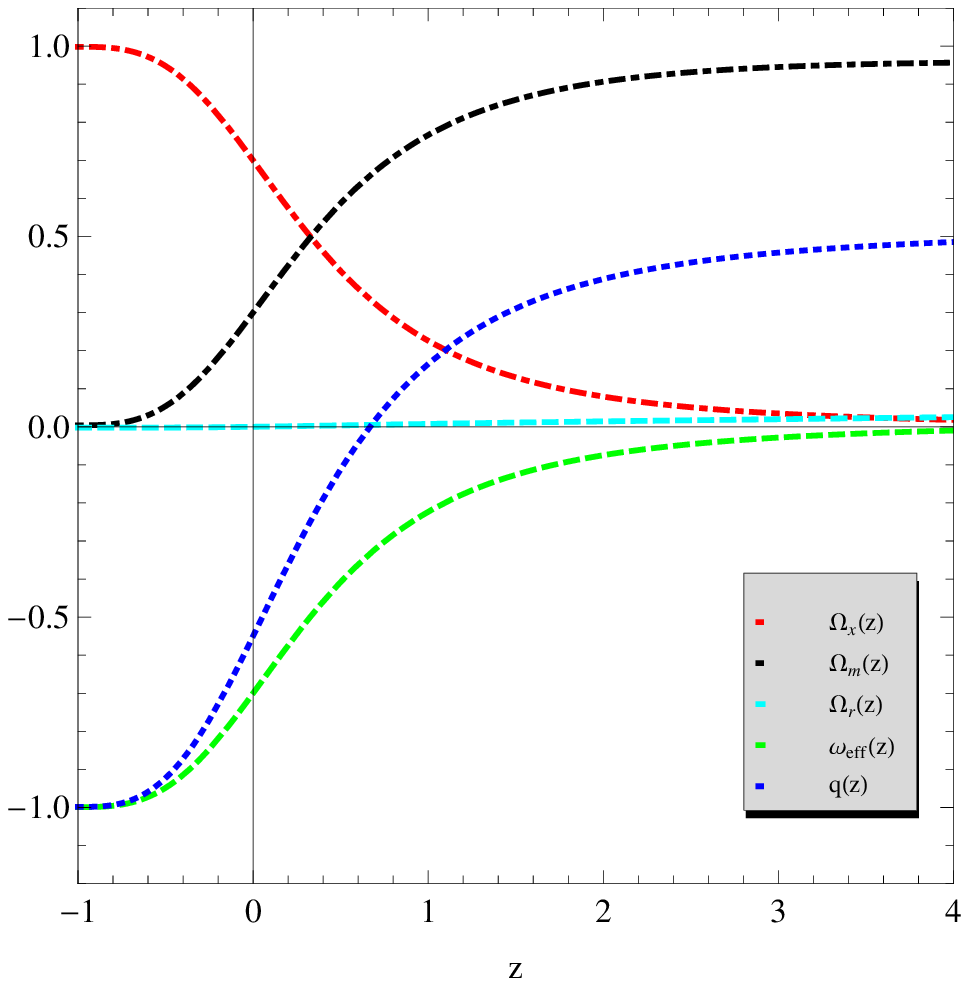}}\hskip0.05cm
\resizebox{1.6in}{!}{\includegraphics{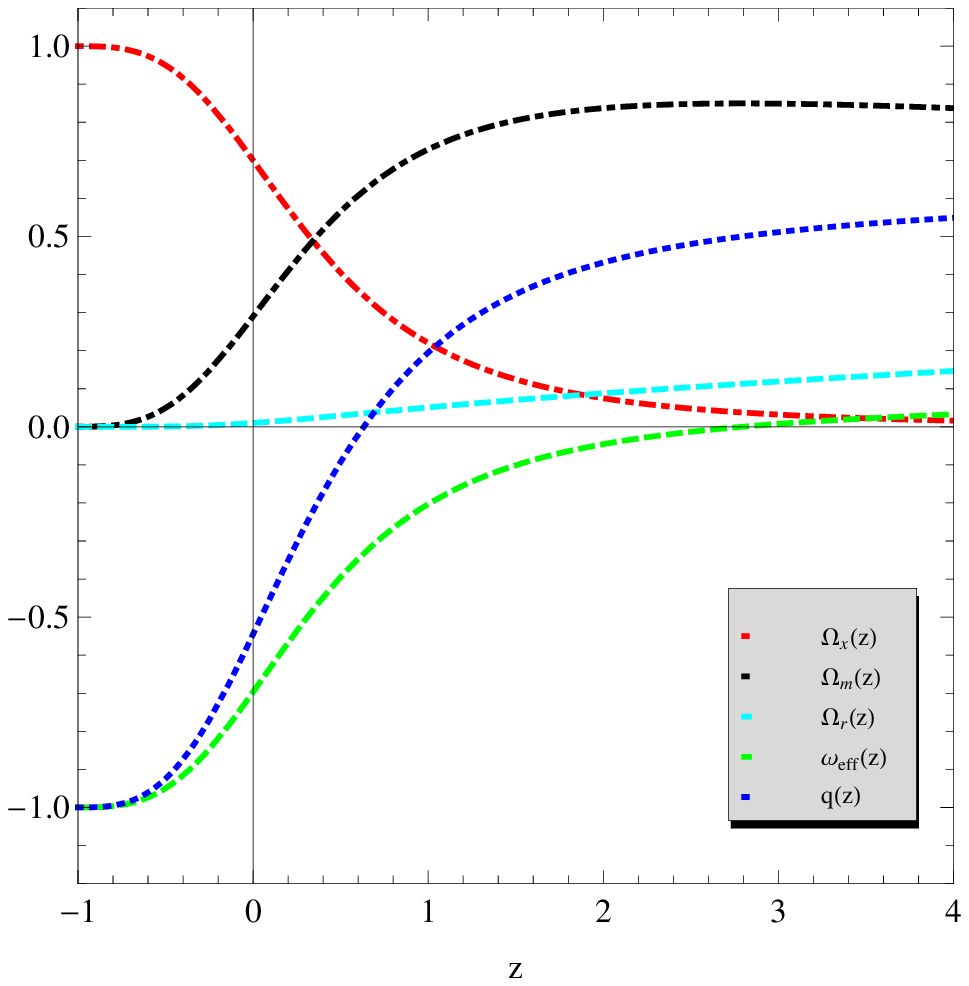}}\hskip0.05cm
\resizebox{1.6in}{!}{\includegraphics{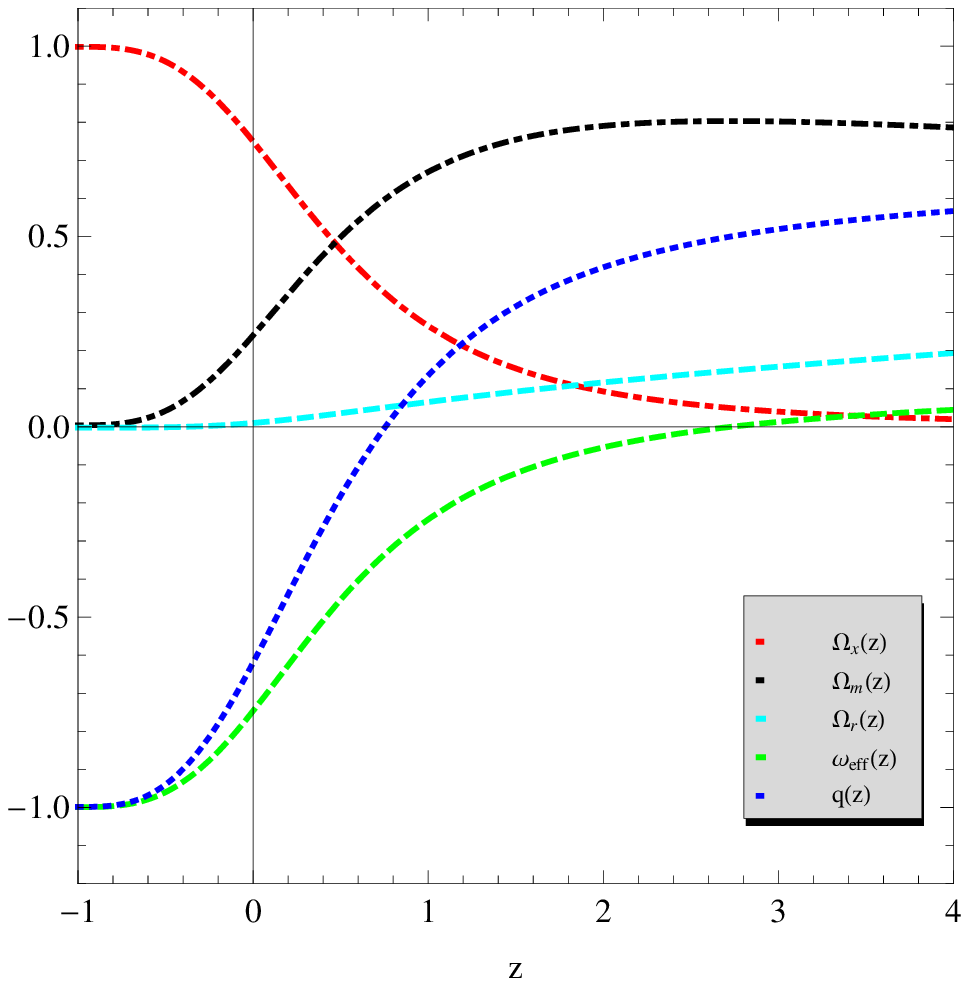}}
\resizebox{1.6in}{!}{\includegraphics{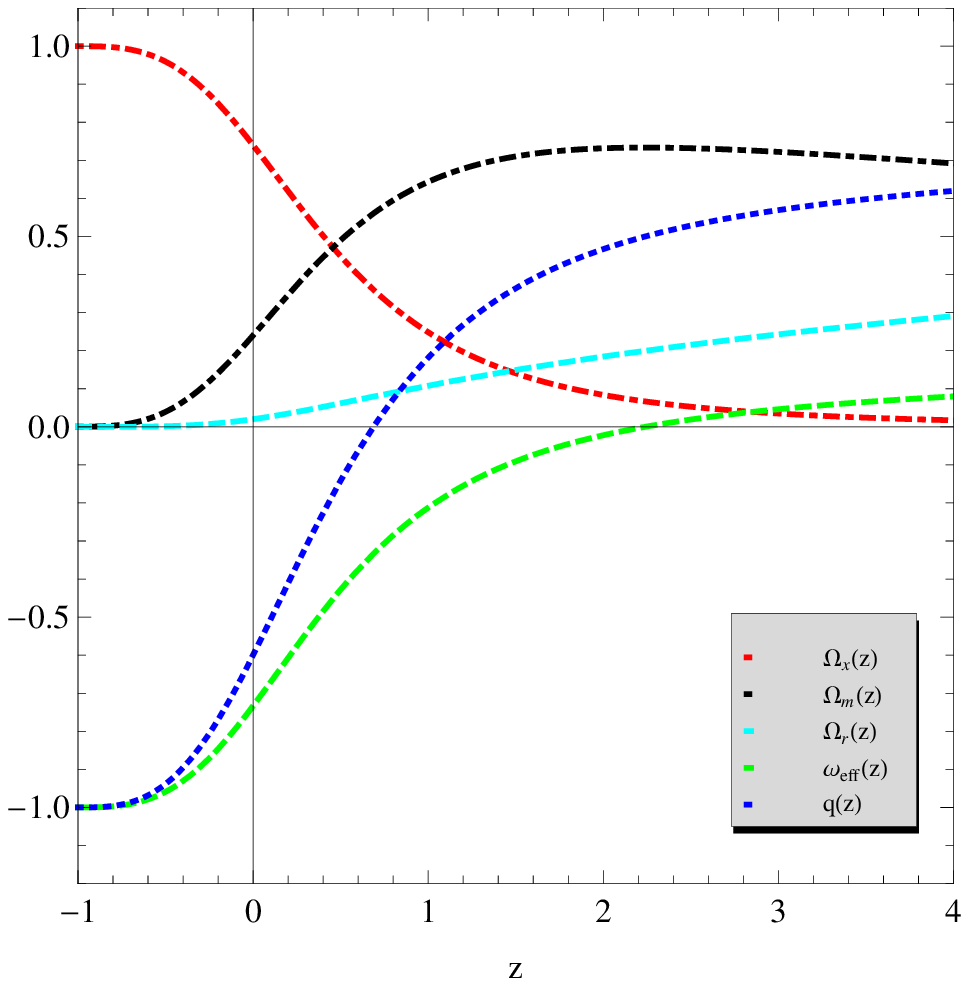}}
\resizebox{1.6in}{!}{\includegraphics{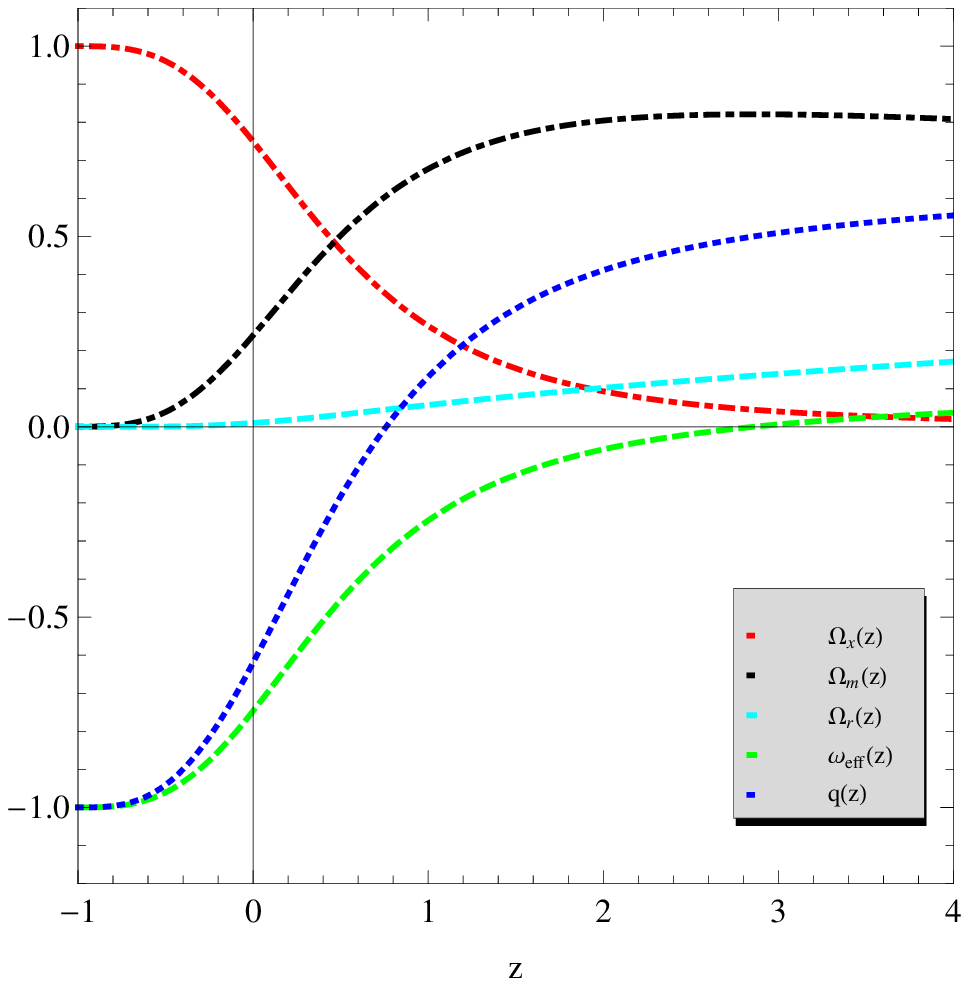}} 
\caption{\scriptsize{Plot of $\Omega_{x}(z)$  $\Omega_{m}(z)$, $\Omega_{r}(z)$, $\omega_{eff}(z)$, and $q(z)$, using the best-fit values  obtained  with the Hubble data for different $\theta$ planes. These plots follows the same  same order used of in two-dimensional C.L. of Fig.1}}
\label{Fig2}
\end{minipage}
\end{figure}

\begin{figure}[hbt!]
\begin{minipage}{1\linewidth}
\resizebox{1.6in}{!}{\includegraphics{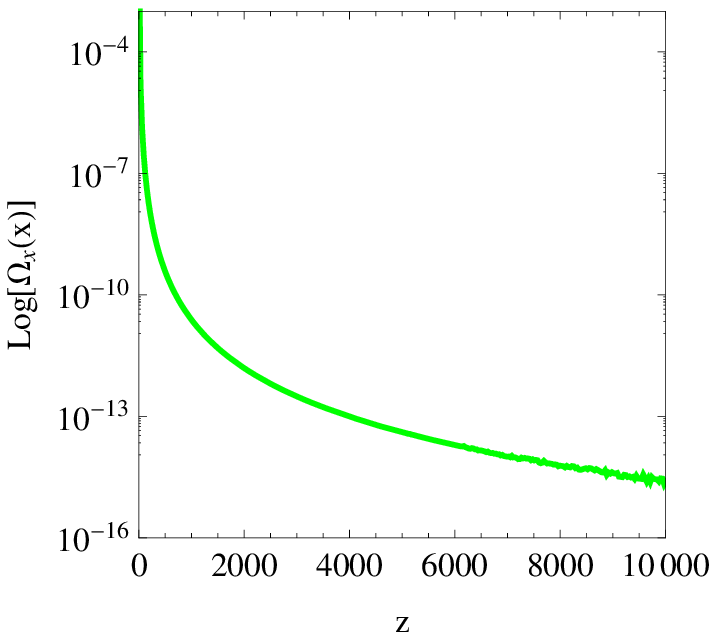}}
\resizebox{1.6in}{!}{\includegraphics{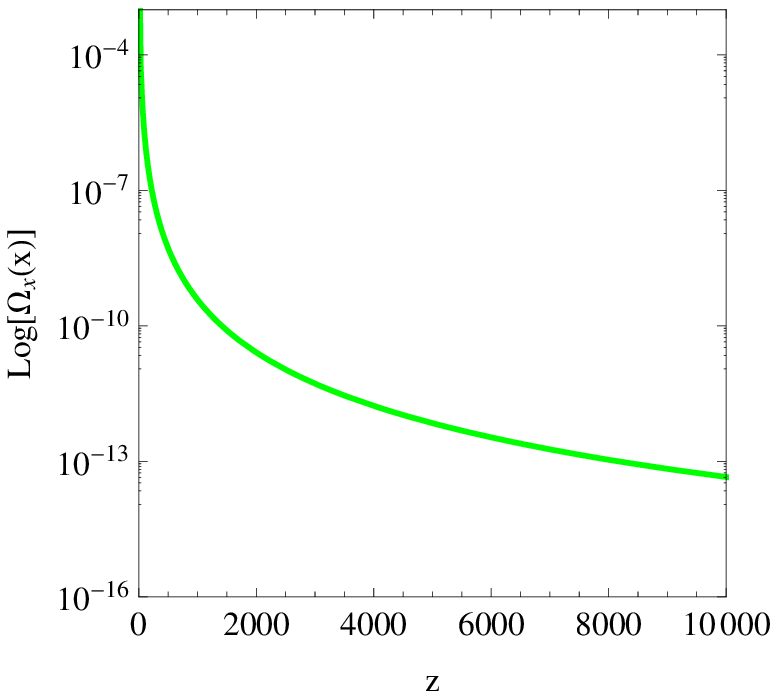}}\hskip0.05cm
\resizebox{1.6in}{!}{\includegraphics{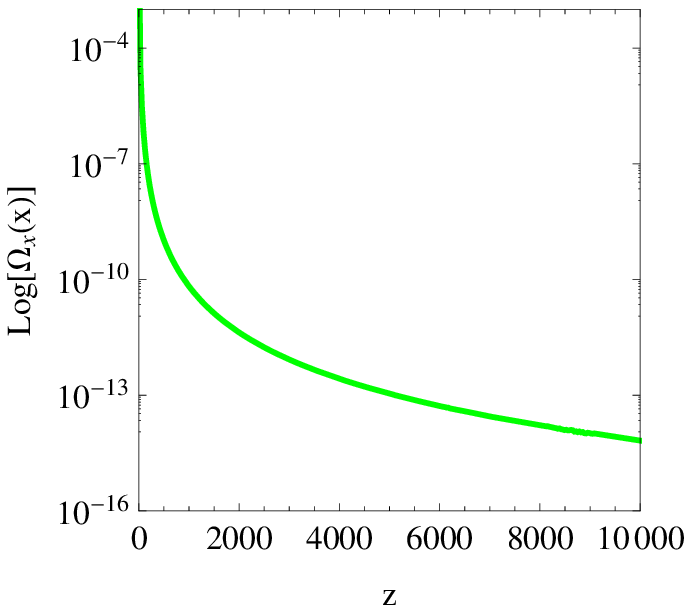}}\hskip0.05cm
\resizebox{1.6in}{!}{\includegraphics{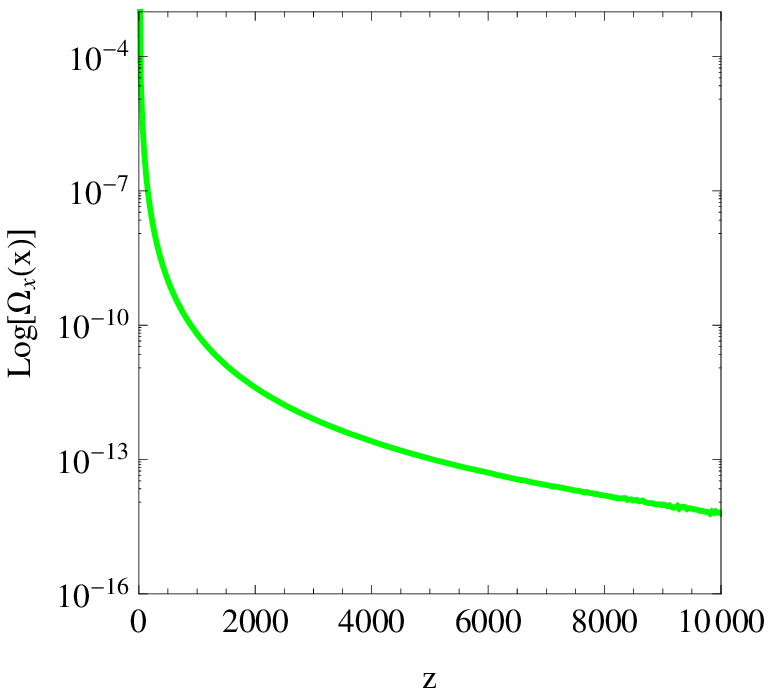}}
\resizebox{1.6in}{!}{\includegraphics{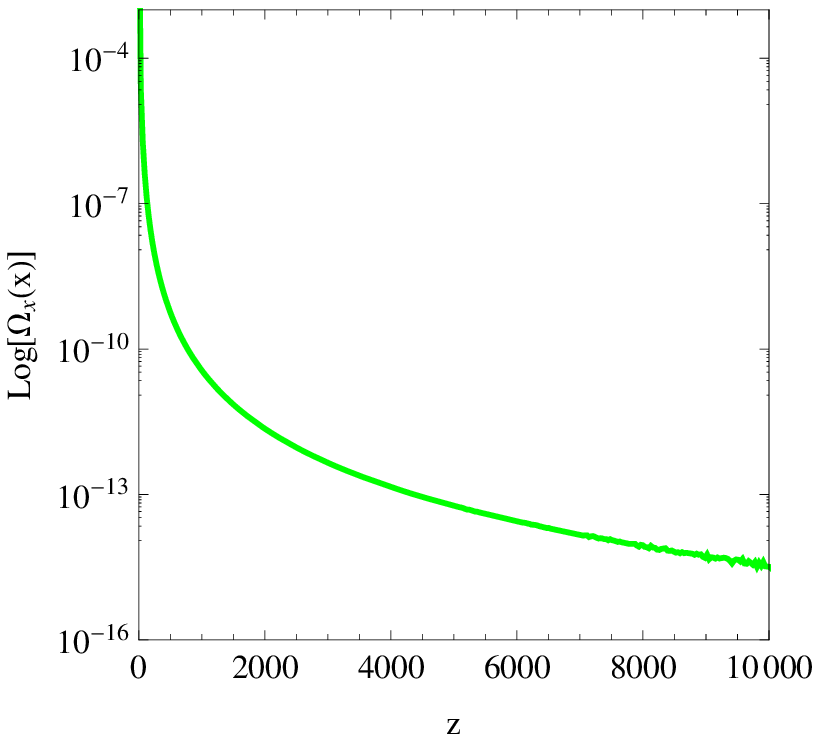}}
\resizebox{1.6in}{!}{\includegraphics{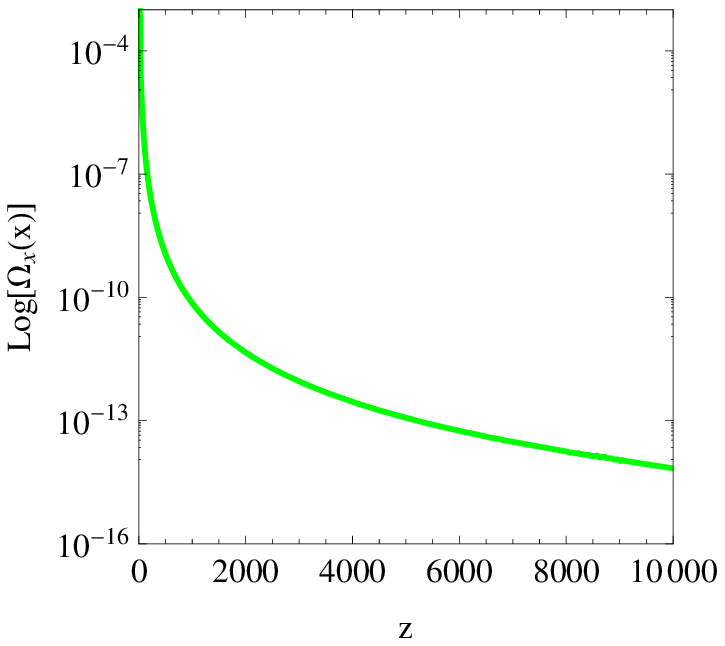}} 
\caption{\scriptsize{Plot of  $\log \Omega_{x} (z)$ for $z \in [10^{2}, 10^{5}]$  using the best-fit values  obtained  with the Hubble data for different $\theta$ planes. These plots follows the same  same order used  in the two-dimensional C.L. of Fig.1}}
\label{Fig3}
\end{minipage}
\end{figure}

The two-dimensional C.L. obtained with the standard $\chi^{2}$ function
for two independent parameters is shown in Fig. (\ref{Fig1}), whereas the estimation
of these cosmic parameters is briefly summarized in Table (\ref{I}). We obtain that
$\ga_{s}$ varies from $10^{-4}$ to $10^{-3}$, so  these values clearly fulfill
the constraint $\ga_{s}<2/3$ that assures the existence of the accelerated phase of
the Universe at late times. We find the best fit at 
$(H_{0}, \Omega_{x0})=(74.32 {\rm km~s^{-1}\,Mpc^{-1}},0.77)$ with $\chi^{2}_{d.o.f}=0.783$
by using the priors $\Omega_{m0}=0.2$ and $\ga_{s}=10^{-3}$. These findings show, in broad terms,  
that the  estimated values of $H_{0}$ and $\Omega_{x0}$ are in agreement with 
the standard ones reported by the WMAP-7 project \cite{WMAP7}. The value of $\Omega_{x0}$
is slightly greater than the standard one of $0.7$ with a discrepancy only of $0.1\%$.
Moreover, we find that using the priors  $(H_{0}, \ga_{s})=(74.20 {\rm km~s^{-1}\,Mpc^{-1}}, 10^{-3})$
the best-fit values for the present-day density parameters are considerably improved, namely, these in  turn give  
$(\Omega_{x0},\Omega_{m0})=(0.74, 0.23)$ along with a lower goodness condition ($\chi^{2}_{d.o.f}=0.779$). 
Regarding the estimated values of $\Omega_{m0}$, we find that it varies from
$0.24$ to $0.29$, without showing a significant difference with the standard ones \cite{WMAP7}. In 
performing the statistical analysis, we find that $H_{0} \in [71.28, 74.32]{\rm km~s^{-1}\,Mpc^{-1}}$ 
so the estimated values are met  within $1 \sigma$ C.L.  reported
by Riess \emph{et al} \cite{H0}, to wit,  $H_{0}=(72.2 \pm 3.6){\rm km~s^{-1}\,Mpc^{-1}}$. 

\begin{center}
\begin{table}
\begin{minipage}{1\linewidth}
\begin{tabular}{|l|l|l|}
\hline
\multicolumn{3}{|c|}{2D Confidence level} \\
\hline
Priors & Best fits & $\chi^{2}_{d.o.f}$\\
\hline
{$(\Omega_{m0},\ga_{s})=(0.2, 10^{-3})$} & $( H_{0},\Omega_{x0} )=( 74.32, 0.77)$& $0.783$\\
\hline
{$(\Omega_{x0}, \ga_{s} )=(0.70, 10^{-3})$} & $( H_{0},\Omega_{m0} )=( 72.28, 0.29)$& $0.834$\\
\hline
{$(\Omega_{x0}, \Omega_{m0})=(0.7, 0.29)$} & $( H_{0},\ga_{s} )=( 71.74, 10^{-4})$& $0.890$\\
\hline
{$( H_{0},\ga_{s} )=( 74.20, 10^{-3})$} & $(\Omega_{x0}, \Omega_{m0})=(0.74, 0.23)$& $0.779$\\
\hline
{$( H_{0},\Omega_{m0} )=( 74.20, 0.24)$} & $(\Omega_{x0},\ga_{s})=(0.74, 10^{-4})$& $0.855$\\
\hline
{$( H_{0},\Omega_{x0} )=( 74.20, 0.75)$} & $(\ga_{s},\Omega_{m0})=(10^{-4}, 0.24)$& $0.776$\\
\hline
\end{tabular}
\caption{\scriptsize{ Observational bounds for the 2D C.L. obtained in  Fig. (\ref{Fig1}) by varying two cosmological parameters.}}
\label{I}
\end{minipage}
\end{table}
\end{center}

\begin{center}
\begin{table}
\begin{minipage}{1\linewidth}
\begin{tabular}{|l|l|l|l|l|l|}
\hline
\multicolumn{6}{|c|}{Bounds for cosmic parameters} \\
\hline
$\theta_{c}$&$z_{t}$ & $q(z=0)$ & $\omega_{eff}(z=0)$ & $\Omega_{r0}$ & $\Omega_{z}(z \simeq 1100)$\\
\hline
{$I$}& $0.75$& $-0.65$& $-0.77$ & $0.02$ & $1.6 \times 10^{-11}$\\
\hline
{$II$}& $0.66$& $-0.55$& $-0.70$& $  10^{-9}$ & $2.6 \times 10^{-10}$\\
\hline
{$III$}&$0.63$& $-0.54$& $-0.69$ & $0.01$ & $4.5 \times 10^{-11}$\\
\hline
{$IV$}&$0.75$& $-0.62$& $-0.74$ & $0.01$& $4.3 \times 10^{-11}$\\
\hline
{$V$}&$0.68$& $-0.60$ & $-0.73$ & $0.02$ & $2.4 \times 10^{-11}$\\
\hline
{$VI$}&$0.75$& $-0.62$ & $-0.74$ & $0.01$ & $4.9 \times 10^{-11}$ \\
\hline
\end{tabular}
\caption{\scriptsize{ Derived  bounds for cosmic parameters using the best fits value of  2D C.L. obtained in  Table. (\ref{I}) by varying two cosmological parameters in six different cases: $I-$ $( H_{0},\Omega_{x0} )=( 74.32, 0.77)$, $II-$ $( H_{0},\Omega_{m0} )=( 72.28, 0.29)$, $III-$ $( H_{0},\ga_{s} )=( 71.74, 10^{-4})$, $IV-$ $(\Omega_{x0}, \Omega_{m0})=(0.74, 0.23)$, $V-$ $(\Omega_{x0},\ga_{s})=(0.74, 10^{-4})$, and $VI-$ $(\ga_{s},\Omega_{m0})=(10^{-4}, 0.24)$.}}
\label{II}
\end{minipage}
\end{table}
\end{center}

For the sake of completeness,
we also report bounds for other cosmological relevant parameters [see Table (\ref{II})], such as the fraction of radiation $\Omega_{r}(z=0)$, the effective  equation
of state at $z=0$ ($\omega_{eff0}=\ga_{eff0}-1$), the decelerating parameter  at the present time $q_{0}$, and the transition redshift ($z_t$); all these quantities are derived
using the six best-fit values reported in Table (\ref{I}). We find that the $z_{t}$ is of the order unity varying over the interval $[0.63, 0.75]$;such values are close 
to $z_{t}=0.69^{+0.20}_{-0.13}$ reported in \cite{Zt1}, \cite{Ztn}  quite recently. Moreover, taking into account
a $\chi^{2}$ statistical analysis made in the $(\omega_{0m}, z_{t})$ plane based on the supernova sample 
(Union2), it has been  shown that at  $2 \sigma$ C.L.  
the transition redshift  varies from $0.60$ to $1.18$ \cite{Zt2}.
The  behavior of a decelerating parameter with redshift is shown in Fig.3, in particular, its present-day
value varies as  $-0.62<q_{0}< -0.54$ for the six cases mentioned in Table II; all these values are in perfectly
agreement with the one reported by the  WMAP-7 project \cite{WMAP7}. In Fig. (\ref{Fig2}) we plot the effective equation of state as a function of
redshit for the best-fit value shown in Table (\ref{II}).  In general,  we find that $-1\leq \omega_{eff}\leq 0$ whereas their
present-day values cover the range $[-0.74, -0.69]$,  so this does not exhibit a quintom phase \cite{NQ}. Regarding the behavior of  density parameters  $\Omega_x$, $\Omega_m$,
and $\Omega_r$, we find that nearly close to $z=0$, the dark energy is the main agent that speeds up the Universe, far away from $z=1$
the Universe is dominated by the dark matter and at very early times the radiation component enter in the action, controlling the entire dynamic
of the Universe around $z \simeq 10^3$[cf. Fig. (\ref{Fig3}) ]. As it was expected the fraction of  radiation at the present moment is negligible; thus, its value varies over  the range  $ 10^{-8}\leq \Omega_{r0} \leq 10^{-3}$ [see Table (\ref{II})].

Now, we seek for another kind of constraint that comes form the physics at early times because this can be considered as a complementary tool for testing
our model. As is well known, the fraction of dark 
energy at the recombination epoch should fulfill the bound $\Omega_{x}(z\simeq 1100)<0.1$ in order for the dark energy model be consistent 
with the big bang nucleosynthesis (BBN) data.   Some signals could arise from the early dark energy (EDE) models, uncovering the nature of DE as well as 
their properties to high redshift, giving an invaluable guide to the physics behind the recent speed up of the Universe \cite{EDE1}. Then,   the current and future data  for constraining the amount  of EDE  was examined, and the cosmological data analyzed has led to an upper bound of  $\Omega_{x}(z\simeq 1100)<0.043$ with $95\%$ confidence level  in the case of relativistic EDE, while  a quintessence type of EDE has given $\Omega_{x}(z\simeq 1100)<0.024$ although   the EDE component is not preferred, it is also not 
excluded from the current data \cite{EDE1}. Another forecast for  the bounds of the EDE  are obtained with the  Planck  and CMBPol experiments\cite{EDE2}; thus,  assuming a $\Omega_{x}(a \simeq 10^{-3}) \simeq 0.03$  for studying 
the stability of this value, it found that $1\sigma$ error coming from the Planck experiment 
is $\sigma^{Planck}_{e} \simeq 0.004$ whereas the CMBPol  improved this bound by a factor of 4 \cite{EDE2}. 

Taking into account the best-fit values reported in Table (\ref{I}), we find that at  early times the dark energy  changes rapidly with the redshift $z$ over the interval 
$[10^{3}, 10^{4}]$ [see Fig. (\ref{Fig3})  for more detail]. Indeed, Table (\ref{II})  shows that it varies as 
$2.4 \times 10^{-11} \leq \Omega_{x}  \leq 2.6 \times 10^{-10}$ around $z \simeq 1100$. Such findings point out that the model constructed here not only fulfills the severe bound
of $\Omega_{x}(z\simeq 1100)<0.1$ but it  is also consistent with the future constraints
achievable by Planck and CMBPol experiments \cite{EDE2} as well, corroborating that the value of the cosmological parameters selected before, 
through the statistical analysis made with Hubble data, are consistent with BBN constraints.

It is interesting to compare the former analysis, 
where radiation was considered as a free evolving component which is decoupled from the dark sector in relation with 
other cases, where a radiation component interacts with 
the dark sector in order to provide a quantitative analysis of 
the role played by the interaction. At present, dark energy dominates the whole dynamics of the Universe and 
there is  practically an obvious decoupling with radiation. 
 However, from a theoretical point of view,  it is reasonable to expect  that dark components can interact with other fluids of the Universe substantially  in the very beginning of its evolution due to a process occurred in the early Universe. For instance, dark energy interacting with neutrinos was investigated in \cite{GK}.
The framework of many interacting components could provide a more  natural  arena for studying  the stringent bounds of dark energy at the recombination epoch.  There could be a signal in favor of  having dark matter exchanging energy with dark energy while  radiation is treated as a decoupled component \cite{hmi1}, \cite{hmi2} or the case where  dark matter, dark energy, and radiation exchange energy. More precisely, when the Universe is filled with an interacting dark sector plus a decoupled radiation term, it was found that $\Omega_{x}(z\simeq 1100)=0.01$ \cite{hmi1} or $\Omega_{x}(z\simeq 1100)=10^{-8}$ \cite{hmi2} but if radiation is coupled to the dark sector, the amount of dark energy is drastically reduced, giving $\Omega_{x}(z\simeq 1100)\simeq {\cal{O}}(10^{-10})$ so the addition of an interacting  radiation term  is important for reducing the amount of early dark energy  in 2  or  8 orders of magnitude.
%%%%%%%%%%%%%%%%%%%%%%%%%%%%%%%%%%%%%%%%%%%%%%%%%%%%%%
\section{conclusion}
%%%%%%%%%%%%%%%%%%%%%%%%%%%%%%%%%%%%%%%%%%%%%%%%%%%%%%%%%%%%%%%%%%%%%%

We have taken under study the dynamical behavior of a cosmological scenario in which the Universe contains  three interacting components, namely, dark energy, dark matter, and radiation within the framework of the usual spatially flat FRW spacetime.    We have worked within the case of a transversal linear interaction because it gives a unique  preferred direction in the plane constraint $\sum_{i}{Q_{i}}=0$ ; then, for such a case we have obtained the partial energy densities  $(\ro_{x}, \ro_{m}, \ro_{r})$  in terms of the total energy density, and its derivatives up to second order.  Additionally, we have imposed  the existence of a power law solution and investigated its stability finding that the system is asymptotically stable in the sense of Lyapunov for $\al>0$ and $\beta>0$ [see Eqs. (\ref{19})-(\ref{19b})]. Using a linear transversal interaction that depends on  the total density and its derivatives up to third order only, we have  found  that the interaction between the three components makes  the total energy density  exhibit three different behaviors with the scale factors $a^{-3\ga_{s}}$, $a^{-3\ga_{+}}$, and $a^{-3\ga_{-}}$, altering the usual corresponding traits to have no interaction. %In order to be able 

On the observational side, we have  examined the previous model  by constraining the cosmological parameters with the Hubble data and the well-known bounds  for dark energy at the recombination era.  In the case of 2D C.L., we have made six statistical constraints with the Hubble function [see Fig. (\ref{Fig2}) and Table (\ref{I})]. We have found that $\ga_{s}$ varies from $10^{-4}$ to $10^{-3}$, so  these values clearly fulfill
the constraint $\ga_{s}<2/3$ for getting an accelerated phase of the Universe at late times.  Using the priors  $(H_{0}, \ga_{s})=(74.20 {\rm km~s^{-1}\,Mpc^{-1}}, 10^{-3})$, the best-fit values for the present-day density parameters are given by $(\Omega_{x0},\Omega_{m0})=(0.74, 0.23)$ along with a $\chi^{2}_{d.o.f.}=0.779<1$. We have obtained that  the estimated values of $\Omega_{m0}$ vary from
$0.24$ to $0.29$, without showing a significant difference with the standard ones \cite{WMAP7}. Besides, it turned  out that  $H_{0} \in [71.28, 74.32]{\rm km~s^{-1}\,Mpc^{-1}}$ so the latter values are met  within $1 \sigma$ C.L.  reported
by Riess \emph{et al} \cite{H0}. Regarding the derived cosmological parameters,  for instance, the transition $z_{t}$ turned to be  of the order unity, varying over the interval $[0.63, 0.75]$; such values  are in agreement with $z_{t}=0.69^{+0.20}_{-0.13}$ reported in \cite{Zt1}-\cite{Ztn} , and meet  within the $2 \sigma$ C.L  obtained with the supernovae (Union 2) data in \cite{Zt2}.  Besides, the decelerating parameter $q(z=0) \in [-0.62, 0.54]$  and  the  equation of state  $\omega_{eff}(z=0) \in [-0.74, -0.69]$; indeed,  $-1\leq \omega_{eff}\leq 0$ [see  Fig. (\ref{Fig3})], while the fraction of radiation at the present momment $\Omega_{r0}$ varies in the interval $ [10^{-3},  10^{-9}]$ for the six cases mentioned in Table (\ref{II}).  Further,  the dark energy amount $\Omega_x(z)$ governs the dynamic of the Universe near $z=0$,  whereas far away from $z=1$ the Universe is dominated by the fraction of dark matter  $\Omega_m(z)$  and at very early times the fraction of radiation $\Omega_r(z)$ controls the entire dynamic of the Universe around $z \simeq 10^3$[cf. Fig. (\ref{Fig3})]. 

Finally, taking into account the best-fit values reported in Table (\ref{I}), we find that at  early times the dark energy  changes rapidly with the redshift $z$ over the interval  $[10^{3}, 10^{4}]$ [cf. Fig. (\ref{Fig3})]; indeed, we have obtained a  $2.4 \times 10^{-11} \leq \Omega_{x}  \leq 2.6 \times 10^{-10}$ around $z \simeq 1100$ [ see Table (\ref{II})]. The latter results indicate that  the model constructed fulfills the severe bound of $\Omega_{x}(z\simeq 1100)<0.1$  and is consistent with the future constraints
achievable by Planck and CMBPol experiments \cite{EDE2} as well, pointing that the cosmological parameters selected before are coherent with BBN constraints also. In a future investigation, we will perform a full study of the interaction vector proportional to orthonormal projection $\mathbf{e_o}$  that is related to the other direction defined in the constraint plane   $\sum_{i}{Q_{i}}=0$.

%%%%%%%%%%%%%%%%%%%%%%%%%%%%%%%%%%%%%%%%%%%%%%%%%%
\acknowledgments
%%%%%%%%%%%%%%%%%%%%%%%%%%%%%%%%%%%%%%%%%%%%%%%%%
We are grateful to the referee for his careful reading of the manuscript.
%We would like to thank the referee for making useful suggestions, which  helped to improve the article.
L.P.C thanks  the University of Buenos Aires under Project No. 20020100100147 and the Consejo Nacional de Investigaciones Cient\'{\i}ficas y T\' ecnicas (CONICET) under Project PIP 114-200801-00328 for the partial support of this work during its different stages. M.G.R is partially supported by CONICET.
%%%%%%%%%%%%%%%%%%%%%%%%%%%%%%%%%%%%%%%%%%%%%%%%%%%%%%%%%%%%%%%%%%%%%%%%%%%%%%%%%%%%%%%%%%%%%%%%%%%%%%%5

%\vspace{1cm}
%%%%%%%%%%%%%%%%%%%%%%%%%%%%%%%%%%%%%%%

\end{document}